\documentclass[pre,preprint,superscriptaddress,nofootinbib]{revtex4-1}

\usepackage{amsmath}
\usepackage{graphicx}

\usepackage{verbatim}

\usepackage{bibunits}
\defaultbibliography{criticalpointjess}
\defaultbibliographystyle{unsrt}

\newcommand{\figref}[1]{Fig.~\ref{#1}}
\newcommand{\sech}{{\text{sech}}}
\renewcommand{\eqref}[1]{Eq.~(\ref{#1})}

\newcommand{\F}{\mathcal{I}}

\newcommand{\hext}{h_{\mathrm{ext}}}

\newcommand{\avg}[1]{\langle #1 \rangle}

\newcommand{\mc}[1]{\ensuremath{\mathcal{#1}}}

\begin{document}

% BCD 6.3.2015 separate bibliography section for main text
\begin{bibunit}

%\baselineskip24pt

%\begin{center}
%\textbf{Control of critical behavior in a small-scale social system}
%\\~\\
%Bryan C. Daniels$^{1}$, David C. Krakauer$^{2}$, Jessica C. Flack$^{2}$
% \\
%% $^1$Center for Complexity \& Collective Computation, Wisconsin Institute for Discovery, Madison, WI\\
%$^1$ASU-SFI Center for Biosocial Complex Systems, Arizona State University, Tempe, AZ\\
%$^2$Santa Fe Institute, Santa Fe, NM
%\end{center}

\title{Control of critical behavior in a small-scale social system}
\author{Bryan C. Daniels}
\affiliation{ASU-SFI Center for Biosocial Complex Systems, Arizona State University, Tempe, AZ}
\author{David C. Krakauer}
\affiliation{Santa Fe Institute, Santa Fe, NM}
\affiliation{ASU-SFI Center for Biosocial Complex Systems, Arizona State University, Tempe, AZ}
\author{Jessica C. Flack}
\affiliation{Santa Fe Institute, Santa Fe, NM}
\affiliation{ASU-SFI Center for Biosocial Complex Systems, Arizona State University, Tempe, AZ}

\begin{abstract}
\baselineskip16pt
Over the last decade new technologies for making large numbers of fine-grained measurements have led to the surprising discovery that many biological systems sit near a critical point \cite{Mora2011, Valverde2015}. These systems are potentially more adaptive in that small changes to component behavior can induce large-scale changes in aggregate structure and function. Examples include networks of neurons \cite{Kastner2015}, ant groups cooperatively carrying a load \cite{GelPinFon15},  and animal groups forming flocks and schools \cite{Bialek2014}.  Accounting for criticality remains a challenge as sensitivity to perturbation suggests a lack of robustness. Furthermore, change induced by perturbation may not be adaptive. Complicating matters further critical phenomena can result from history-dependent stochastic processes \cite{Corominas2014}. A question central to distinguishing among these conflicting views of criticality is to what degree criticality can be controlled by the components of the system \cite{Valverde2015}. We address the control of criticality using data on conflict dynamics and fight sizes from an animal society model system (\emph{Macaca nemestrina}, $n$=48). The system is fundamentally finite so we operationalize criticality in information theoretic terms using Fisher information and a measure of  instability. We analyze criticality using empirically-grounded equilibrium (maximum entropy) and dynamic (branching process) models of the monkeys' fight-joining behavior. We find that (1) this \emph{heterogeneous, socially} organized system, like homogeneous, spatial systems (flocks and schools), sits near a critical point, (2) the contributions individuals make to how critical the system is can be quantified and vary, and (3) the distance from the critical point (DFC) can be controlled through biologically plausible mechanisms operating on this heterogeneity. These mechanisms include third-party policing, which dampens fight participation of the individuals with the largest effect on DFC \cite{Flack:2005ih,Flack:2006vi,Flack:2005dg}. Control of DFC allows biological systems to balance the tradeoff between robustness and need for rapid change.
\end{abstract}

\maketitle

%\section*{}
We analyze a time-series of fights from a large, captive pigtailed macaque group collected over multiple observation periods during a four month period (SI Sec. \ref{EmpiricalMethods}). The data consist of a series of binary fight participation vectors $\vec x$ of length $n$. For each vector an individual is assigned a ``1'' if it participated in that fight and a ``0'' if it did not (SI, Sec. \ref{EmpiricalMethods}).

To study criticality in our model system we can use tools from statistical mechanics. These tools are best deployed when the study system can be described within a simple, tractable, and well understood modeling framework. Hence our first task is to assess whether our data are consistent with any of three basic, but biologically valid, fight joining models: (1) independent fight joining decisions, (2) correlated fight joining decisions  (equilibrium model), and (3) strategic, correlated fight joining decisions (dynamical branching process) (SI, Sec. \ref{modeldescription}) \cite{Krakauer:2011tb}.  We evaluate these models by determining how effectively each recovers a key social feature --- the distribution of fight sizes $s$ \cite{Dedeo2010} (SI, Sec.\ref{modeldescription})---when parameterized by the empirical data.

In the independent model individuals join fights without taking into account which others are currently fighting (SI, Sec. \ref{modelevaluation}). 

The correlated decision-making model is an equilibrium maximum entropy model that fits all pairwise correlations and corresponds to a spin-glass model \cite{DanKraFla12}. The resulting probability distribution over possible fights has relative negative log-likelihood,
\begin{equation}
\label{isingModel}
L(\vec x) = -\sum x_i J_{ij} x_j,
\end{equation}
where the coefficients $J_{ij}$ are numerically fit to match individual and pairwise frequencies $\avg{x_i}$ and $\avg{x_i x_j}$.

In the dynamical branching process the dominant causes of conflict are temporal pairwise interactions---an individual joins the current fight with some finite probability only when it sees another individual join. Fight initiation is assumed to occur at a slower timescale, when a random individual becomes aggressive. Individuals join the fight by receiving aggression from or initiating aggression against an individual already in the fight. Parameters include individual initiation parameters $p_{0i}$, each denoting the relative probability that individual $i$ begins a fight, and conditional redirection parameters $p_{ij}$, each denoting the probability that $j$ joins a fight due to $i$ having just joined.  We consider only the relative ordering of individuals joining fights, not direct interactions, and estimate parameters by fitting conditional frequencies $P_{ij}$ --- the frequency of seeing individual $j$ at any later time in a fight sequence given that $i$ appeared first.

% $P_{ij} = N_{ij}/N_{i}$

\begin{figure}
\centering
\includegraphics[scale=0.85]{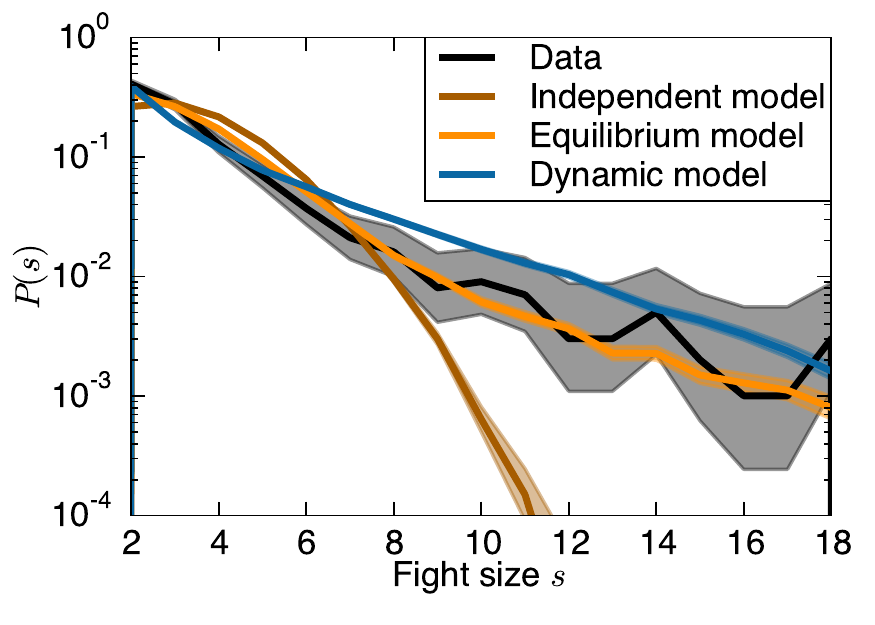}
\caption{ Testing fight-joining processes fit to individual-level data:
    Models that include social correlations (the equilibrium maximum entropy and dynamic branching process models) can reproduce the relatively long tail of the observed distribution of fight sizes, whereas assuming independent fight joining events (the independent model) cannot.
\label{fightSizeDistributions}%
}
\end{figure}

We test the performance of each model by its ability to predict the distribution of fight sizes (\figref{fightSizeDistributions}) and correlations up to 3rd order (\figref{isingStatistics}) (SI, Sec. \ref{modelevaluation}). We rule out the independent model (SI, Sec. \ref{modelevaluation}).  We find the empirically parameterized maximum entropy (as in prior work \cite{DanKraFla12}) and branching process models recover the observed distribution of fight sizes, indicating that these models are mechanistically consistent with the data. Because these models are well understood modeling frameworks in statistical physics, they offer a rigorous foundation for proceeding with a statistical mechanical description of the system. This allows us to define and quantify the sensitivity to perturbations and characterize criticality in a finite system.

A phase transition (SI, Sec~\ref{evalsens}) can be thought of as sensitivity that diverges in the limit of infinite number of individuals due to a collective instability. This instability is an aggregate-level property that causes perturbations to individuals to be amplified and spread to change the behavior of the entire system (SI, Sec~\ref{collectinstab}).  While phase transitions are typically identified by examining the asymptotic behavior of the infinite limit, social systems are often not well approximated by such a ``thermodynamic'' limit.  Yet sensitivity and instability are still important concepts in these systems.  Here, in both the maximum entropy and dynamical branching models of our finite system, we find that changes to key parameters can lead quickly to increased sensitivity and instability.  We take this to be an operational definition of a ``critical point'' in a finite system (for a review of other definitions, see \cite{Valverde2015}).

We operationalize sensitivity as the derivative of average fight size with respect to an individual's agitation, averaged over individuals.  In the equilibrium model, the control parameter associated with the average fight size $\avg{s}$ is an external ``field'' $\hext$, which we interpret as uniformly increasing each individual's agitation, making it more likely to become involved in fights. Adding an external field to \eqref{isingModel}, we have $L(\vec x) = -\sum x_i J_{ij} x_j - \hext \sum x_i$. The corresponding sensitivity is equal to the susceptibility $\chi$:
\begin{equation}
\frac{1}{n} \sum_i{\frac{d \avg{s}}{d h_i}}
 = \chi = \frac{1}{n} \frac{d \avg{s}}{d \hext}.
\end{equation}

\begin{figure}
\centering
\includegraphics[scale=0.85]{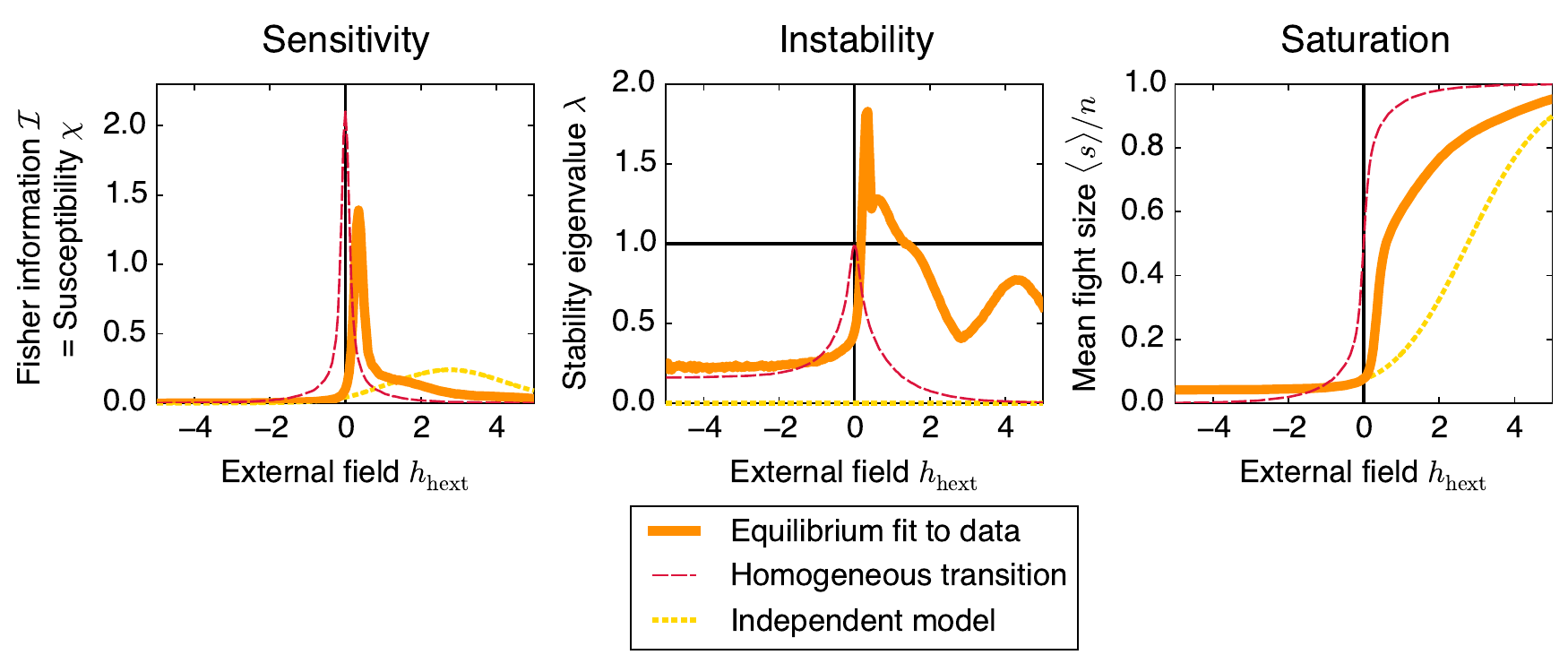}
\caption{ \label{susceptibilityFigure}%
The system is near a sensitive and unstable region of parameter space. (Left) The susceptibility $\chi$, measuring the sensitivity of average fight size to an external perturbation, has a peak near the fit parameters (at $\hext = 0$). The susceptibility is also equal to the Fisher information $F(\hext)$, describing how quickly the distribution over fights becomes distinguishable as $\hext$ is varied. (Middle) Instability of the lowest order mean-field solution is indicated by the eigenvalue $\lambda$ becoming larger than 1. Yellow dashed lines indicate a system of the same size with independent individuals, and red dashed lines indicate a homogeneous system of the same size tuned to be marginally unstable at $\hext = 0$.  (Right) The transition is associated with large changes in mean fight size.
}
\end{figure}

Results for the equilibrium model are shown in \figref{susceptibilityFigure}A. For fit parameters ($\hext = 0$), there is increased sensitivity compared to a noninteracting model with the same mean fight size, meaning that an amplification process is occurring that makes changes to patterns of aggression at the individual level ``visible'' at the global system level. This amplification process cannot be attributed to external events as these data were collected in a captive setting in which such disturbances were minimized (SI, Sec~\ref{modeldescription}).

The susceptibility can also be interpreted as a Fisher information, an information theoretic quantity that describes how sensitive a distribution is to the parameters that describe it (SI~\ref{FisherInfoSection}). A large $\chi$ implies faster learning: large susceptibility means that aggregate level statistics are informative about conflict dynamics at the individual level \cite{Tchernookov2012,Prokopenko2011}.

Analogously to the susceptibility in the equilibrium model, we can define sensitivity in the dynamic model as how quickly fight sizes grow following perturbation of redirection probabilities $p_{ij}$:
\begin{equation}
    \frac{1}{n(n-1)}\sum_i \sum_{j \neq i} \frac{d\avg{s}}{dp_{ij}}
    \equiv \chi_\mathrm{dyn}
    = \frac{1}{n(n-1)} \frac{ d\avg{s} }{ dp },
\end{equation}
where $p$ adds probability uniformly to all redirection probabilities.

\begin{figure}
\centering
\includegraphics[scale=0.85]{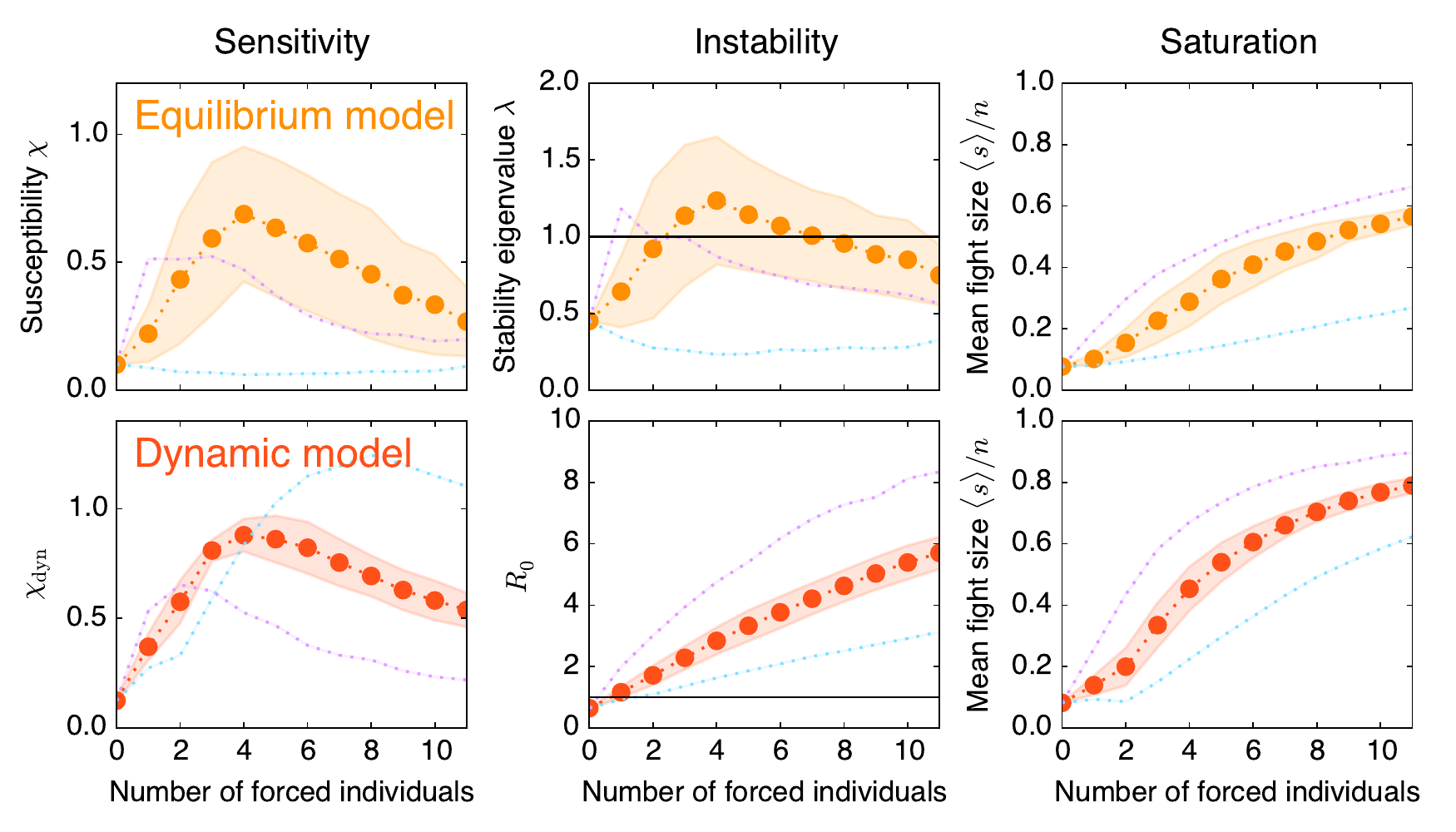}
\caption{ 
    In equilibrium (top) and dynamical (bottom) models,
    forcing a few individuals to become simultaneously
    aggressive leads to a more sensitive system. (Left)
    In each model, peak sensitivity is reached when on average
    3-5 individuals are forced. Shaded regions indicate
    the standard deviation around the mean over different
    realizations of the chosen individuals. Purple and
    blue dotted lines indicate choices of forced individuals
    that maximize and minimize the resulting mean fight size.
    (Middle) The mean fight size.
    (Right) The lowest-order stability
    measured in each model, corresponding to the
    eigenvalue $\lambda_0$ in the equilibrium model
    and $R_0$ in the dynamical model (SI, Sec~\ref{collectinstab}).  A value above
    1 in each case indicates instability to linear order.
\label{ellStarFigure}%
}
\end{figure}

\figref{susceptibilityFigure} suggests that the system is near a transition. To quantify this statement in a biologically meaningful way we measure how sensitivity and stability vary as we force---through parameterized-simulation---some number of individuals to fight continuously (while allowing the system to consist only of the non-forced individuals). The number of individuals required to reach a peak in sensitivity corresponding to a collective instability (SI, Sec.\ref{collectinstab}) is a measure of the \emph{distance of the system from the critical point} (DFC). Beyond this peak lies saturation and decreasing sensitivity.

In the equilibrium model forcing is accomplished by removing the forced individuals and adding their interaction terms to the fields acting on the remaining individuals: the new fields are given by $J^*_{ii} = J_{ii} + \sum_{\mathrm{forced}~k} 2 J_{ik}$. In the dynamic model this is accomplished by explicitly including the forced individuals at each timestep in the simulation but only recording the behavior of the remaining individuals.  The change in sensitivity caused by forcing each individual also provides ``sensitivity scores'' that measure the extent to which each individual brings the system closer to criticality.
%Because most significant interactions are excitatory we expect these forced
%individuals to cause an increase in the mean fight size of the remaining individuals.

As shown in \figref{ellStarFigure}, both models predict a substantial increase in sensitivity when a few individuals are forced to be continuously involved in conflict. The system begins to saturate in sensitivity after 3-5 individuals are simultaneously forced. This result is similar to the finding that bird flocks require only a small proportion of informed individuals to correctly choose direction\cite{COUZIN2005,LEONARD2012}. 

We use perturbation theory to demonstrate that this sensitivity arises from instability in each model (SI, Sec.\ref{stabilitySection}). The number of individuals required to reach a peak in sensitivity corresponding to an aggregate-level instability (SI, Sec.\ref{collectinstab}) is a \emph{biologically meaningful measure} of the \emph{distance of the system from the critical point}.

In addition we find that individuals vary in their sensitivity scores such that forcing participation of individuals with the top two to three sensitivity scores is sufficient to move the system close to a critical point (SI, Sec.\ref{sensitiveindividuals}).
% [whereas decreasing involvement is sufficient to move the system away, towards robustness]

These results tell us that fight involvement by the high sensitivity score individuals can be tuned two ways: through \emph{direct tuning}, as they opportunistically or strategically increase or decrease their fight participation, and \emph{indirect tuning}, through the actions of others that increase or decrease the frequency with which aggression is directed or redirected at these individuals (SI, Sec.\ref{EmpiricalMethods} and SI, Secs.\ref{tuning}). This includes through the policing mechanism in which powerful third parties to the fight intervene and impartially break it up  (SI, Sec.\ref{EmpiricalMethods} and SI, Secs.\ref{tuning}) \cite{Flack:2005dg}. Policing dampens fight involvement by significantly reducing how frequently fights erupt and also by reducing the frequency of redirected aggression received, particularly by high sensitivity individuals\cite{Flack:2005ih}. Hence DFC can in principle be controlled through policing or by the high sensitivity individuals themselves as they up or down regulate the frequency with which they join fights. 

All biological systems need to balance stability and robustness with the need for rapid adaptive change. Yet many biological systems are observed to sit near a critical point, which would seem to suggest a lack of robustness. This apparent conflict can be resolved by the discovery that DFC can be tuned through realistic social mechanisms. Tuning DFC allows for switching between stability and criticality, providing a means for accessing alternative social structures that may be more appropriate if and when the environment should change\cite{FlaErwEll13}. This discovery raises many new questions: It is one thing for tuning to be possible and another to know when to tune and whether to tune towards robustness or criticality (SI, Sec.\ref{adaptive value}). We propose robustness is a good strategy when the environment is stable and low variance. Criticality is a good strategy when the environment is uncertain (SI, Sec.\ref{adaptive value}). In addition for tuning to be adaptive the state of the environment must be accurately perceived by, or mirrored in, the individuals whose behavior changes DFC. This appears to be the case in our system (SI, Sec.\ref{adaptive value}) but may present a crucial evolutionary challenge for other systems.  Future comparative studies are required to quantify the range of DFC across social groups within a species, as well as across biological systems more generally, and to study how DFC might be controlled in other systems. We can then assess whether variation in DFC and its control are correlated with rate of change in the environment and or environmental uncertainty as we expect.

%\begin{thebibliography}{11}

%\bibliography{criticalpointjess.bib}

\section*{End Notes}\begin{itemize}
%\item Supplementary Information is linked to the online version of the paper at www.nature.com/nature.
\item The authors thank Chris Ellison, Philip Poon, Eddie Lee, Eleanor Brush, Yoav Kallus, and Raissa D'Souza for helpful discussion. Partial support for this project was provided by the Templeton Foundation through grants to study complexity and the mind-brain problem, and by the Office of Army Research under contract W911NF-13-1-0340. JCF thanks Frans de Waal and the animal care staff at Yerkes National Primate Research Center for support during data collection.
\item BCD designed the study, built the models, analyzed and interpreted data, and wrote the paper. DCK contributed to model development, analysis, and interpretation, and wrote the paper. JCF designed the study, collected the data, contributed to model development, analysis, and interpretation, and wrote the paper.
Correspondence and requests for materials should be addressed to Bryan Daniels (bryan.daniels.1@asu.edu).
%\item Reprints and permissions information are available at www.nature.com/reprints.
%\item The authors have no competing interests.
\end{itemize}

% BCD 6.3.2015 new bibliography unit for SI
\putbib
\end{bibunit}

\begin{bibunit}

\clearpage

% change citation format in Supplementary Material
\renewcommand*{\citenumfont}[1]{S#1}
\renewcommand*{\bibnumfmt}[1]{[S#1]}

\begin{center}
\textbf{Supplementary Information}
\\~\\
Control of critical behavior in a small-scale social system
\\
Bryan C. Daniels, David C. Krakauer, Jessica C. Flack
\end{center}

\section{Empirical Methods}
\label{EmpiricalMethods}
\subsection{Ethics statement}
The data collection protocol was approved by the Emory University Institutional Animal Care and Use Committee and all data were collected in accordance with its guidelines for the ethical treatment of nonhuman study subjects.

\subsection{Study system}
The data were collected by JCF in 1998 from a large group of captive pigtailed macaques (\emph{Macaca nemestrina}) socially-housed at the Yerkes National Primate Center in Lawrenceville, Georgia. Pigtailed macaques are indigenous to south East Asia and live in multi-male, multi-female societies characterized by female matrilines and male group transfer upon onset of puberty~\cite{Caldecott:1986uk}. Pigtailed macaques breed all year. Females develop swellings when in \OE strus. Macaque societies more generally are characterized by social learning at the individual level, social structures that arise from nonlinear processes and feed back to influence individual behavior, frequent non-kin interactions and multiplayer conflict interactions, the cost and benefits of which can be quantified at the individual and social network levels~\cite {Flack:2007ir, Flack:2006jh, Flack:2006vi, Flack:2005ih, Flack:2005dg, Thierry:2004tj}.

The study group contained $n = 48$ socially-mature individuals (we exclude non-mature individuals because their behavioral strategies are still developing and so are non-stationary over short timescales) and 84 individuals in total. Socially-mature males were at least 48 months and socially-mature females were at least 36 months by study start. These thresholds correspond to approximate onset of social maturity in pigtailed macaques. The study group had a demographic structure approximating wild populations and subadult and adult males were regularly removed to mimic emigration occurring in wild populations.  All individuals, except 8 (4 males, 4 females), were either natal to the group or had been in the group since formation. The group was housed in an indoor-outdoor facility, the outdoor compound of which was 125 x 65 ft.

Pigtailed macaques have frequent conflict and employ targeted intervention and repair strategies for managing conflict \cite{Flack:2005dg}. Data on social dynamics and conflict were collected from this group over a stable, four month period. Operational definitions are provided below in SI Sec.\ref{op}.

\subsection{Operational definitions}
\label{op}
\emph{Fight}: includes any interaction in which one individual threatens or aggresses a second individual. A conflict was considered terminated if no aggression or withdrawal response (fleeing, crouching, screaming, running away, submission signals) was exhibited by any of the conflict  participants for \emph{two minutes} from the last such event. A fight can involve multiple individuals. Third parties can become involved in pair-wise conflict through intervention or redirection, or when a family member of a conflict participant attacks a fourth-party. Fights in this data set ranged in size from 2 to 31 individuals, counting only the socially-mature animals. Fights can be represented as small networks that grow and shrink as pair-wise and triadic interactions become active or terminate until there are no more individuals fighting under the above described two minute criterion. In addition to aggressors, a conflict can include individuals who show no aggression or submission (\emph{e.g.} third-parties who simply approach the conflict or show affiliative / submissive behavior upon approaching, and recipients of aggression who show no response to aggression (typically, threats) by another individual). Because conflicts involve multiple actors, two or more individuals can participate in the same conflict but not interact directly.

In this study only information about fight composition (which individuals were involved) is used.  Only fights that included two or more socially-mature individuals were used in the analysis; the data includes $N = 994$ such fights.  We do not consider internal aspects of the fight, such as who does what to whom, except for the order of each individual's first involvement in the fight (used to estimate time-ordered conditional probabilities for use in the dynamical branching process model). Time data were collected on fight onset and termination but are not used in these analyses.

\emph{Power}: The degree of consensus among individuals in the group about whether an individual is capable of using force successfully~\cite{Flack:2006jh,Brush2013}. In previous works we showed that consensus can be quantified by taking into account the total number of subordination signals an individual receives and multiplying this quantity by a measure of the diversity of signals received from its population of signalers (quantified by computing the Shannon entropy of the vector of signals received by individual \emph{i}) \cite{Flack:2006jh}. In pigtailed macaque societies, the subordination signal is the silent bared teeth display~\cite{Flack:2007ir} emitted outside the conflict context during pass-byes and affiliative interactions. The distribution of power in our study group is heavy tailed, such that a few individuals are disproportionately powerful.

\emph{Policing}: A policing intervention is an impartial intervention performed by a third party into an ongoing conflict~\cite{Flack:2005dg}. Three males and one female perform the majority of effective policing interventions but only the three males (Eo, Qs, Fo) specialize on policing~\cite{Krakauer:2011tb}. These four individuals occupy the top four spots in the power distribution~\cite{Flack:2005dg, Brush2013}.

\emph{Redirection}: A redirection occurs when an aggressor, recipient, or intervener directs aggression at a third (or fourth) individual who was not its original target or attacker. The target of the redirection may not have been involved in the fight until the redirection, or may have been involved in the fight but interacting with individuals other than the redirecting individual. 

\subsection{Data collection protocol}
\label{DataCollectionSection}
The data were collected by a trained observer (J.C.~Flack). The observer spent roughly 100 hours prior to data collection learning to recognize individuals and accurately code their behavior from the observation tower above the monkey compound. Accuracy was validated by a second trained observer (F.B.M.~de Waal). JCF also evaluated coding accuracy using video. Coding accuracy was nearly 100$\%$. 

During observations all individuals were confined to the outdoor portion of the compound and were visible to the observer. The $\approx150$ hours of observations occurred for up to eight hours daily between 1,100 and 2,000 hours over a twenty-week period from June through October 1998, and were evenly distributed over the day. Conflict and signaling data were collected using all-occurrence sampling in which the entire conflict event is followed from start to finish and all participants and their behavior are recorded.

Provisioning occurred before observations and once during observations at approximately the same time each day. The group was stable during the data collection period (defined as no reversals in status signaling interactions resulting in a change to an individual's power score; see Ref.~\cite{Flack:2006jh}). 

\section{Notes on model descriptions and justification}
\label{modeldescription}
We first evaluate which of three basic, empirically-grounded fight joining models explain our macroscopic observable, the distribution of fight sizes. We only accept a model if it is both consistent with the measured, microscopic data and can recover in simulation the observed, measured macroscopic output. Hence all of our models are closely tethered to the measured data and biological details of our model system.

It is important to realize that the parameters in the maximum entropy and branching models come from the microscopic data. The models do not assume prescribed values for parameters but are perhaps better viewed as hypotheses about the ways in which the measured, microscopic detail is connected to observed macroscopic patterns.

All models assume that events external to the system do not create correlations in behavior.  The data were collected in a controlled, captive setting designed to minimize the influence of external events, and we have no evidence for important, consistent external forcers of conflict.

\subsection{Independent model description}The independent model assumes individuals do not respond to each other but instead join fights without regard to who else is fighting.  In this case, perturbations to individual conflict behavior would have no additional effect on group behavior.

The independent model fails to recover both the observed distribution of fight sizes (\figref{fightSizeDistributions}) and the observed significant pairwise correlations (\figref{isingStatistics}).

\subsection{Pairwise maximum entropy model description}Individuals sometimes randomly join fights and sometimes the decision to fight reflects strategic interactions at a pairwise level\cite{Dedeo2010}. We capture this interpretation of the microscopic data using a maximum entropy approach, which corresponds to the spin-glass model of magnetic systems in physics.
Because the model is parameterized by the data it is empirically grounded and serves as a valid biological hypothesis.  However we note that it is less mechanistically specific than the branching process described below: spin-glass interactions are not directional or time-ordered, but rather operate symmetrically and over the timescale of an entire fight bout.

The pairwise maximum entropy model, with parameters fit from the microscopic data, recovers well the distribution of fight sizes (\figref{fightSizeDistributions}). The good performance of the model leads to the prediction that the sensitivity $\chi$ is about twice that of a system with the same conflict frequencies but no strategic interactions.

\subsection{Branching process description}Another reasonable interpretation of the microscopic data is that the random component of fight joining decisions is very small and the decision to fight reflects strategic interactions at a pairwise level. This leads to a branching process model that, with parameters fit from the microscopic data, also recovers the observed distribution of fight sizes. The collective behavior produced by this model can be simply understood in terms of a single parameter, the branching ratio. Additionally, branching process models like this one have been used in previous work to explore other aspects of conflict dynamics in this system, including the role of policing and other forms of third-party intervention in the infectivity of aggression \cite{Krakauer:2011tb}.

\section{Notes on model inference}

\subsection{Independent model inference}
\label{independentInference}
The independent model consists of individuals participating in conflict randomly,  with the frequencies of individual appearance equal to their empirically measured, heterogeneous values $f_i = \avg{x_i}$.  Naively, this can be written as a relative negative log-likelihood\footnote{The relative negative log-likelihood $L(x)$ of state $x$ is related to the likelihood $p(x)$ by $p(x) = \exp(-L(x))/Z$, where $Z$ is a normalization constant set by the constraint that the sum of likelihoods over all states is one.  In statistical physics, $Z$ is the partition function and $L(x)$ is proportional to the free energy of state $x$.} $L^{\mathrm{ind}}(x) = \sum_{i} - h_i x_{i}$, and is equivalent to a maximum entropy model that matches only the frequencies $f_i$.

However, as detailed in SI Sec~\ref{op}, a fight was operationalized for these analyses as involving two or more
socially-mature individuals. (Observed fights that involved only juveniles and 0 or 1 mature individuals are therefore excluded.)
Correspondingly, we forbid our models from producing fights of size smaller than two. We treat this as an additional constraint on the model. The resulting maximum entropy model is then the one in which the likelihood of states with fewer than two individuals present is taken to zero.  This corresponds to a relative negative log-likelihood
\begin{eqnarray}
\label{constrainedIndependentEnergy}
L^{\mathrm{ind}}_{\alpha \rightarrow \infty} &=&
\lim_{\alpha \rightarrow \infty} \left[
\sum_i h_i x_i
+ \alpha ~ \Theta \left( 2 - \sum_j x_j \right) \right],
\end{eqnarray}
where $\Theta(z)$ is 0 when $z \leq 0$ and 1 when $z > 0$.

In the unconstrained case ($\alpha = 0$), we can easily solve for $h_i$:
\begin{eqnarray}
\label{unconstrainedIndependentEnergy}
L^{\mathrm{ind}}_{\alpha = 0} &=& \sum_i h_i x_i                         \\
\label{fieldsEqn}
h_i &=& - \log \left( \frac{\avg{x_i}_{\alpha = 0}}
{1 - \avg{x_i}_{\alpha = 0}} \right)    .
\end{eqnarray}
We must now solve numerically for $h_i$ to match
the empirically measured $f_i = \avg{x_i} = \avg{x_i}_{\alpha \rightarrow \infty}$.
To accomplish this, note that the unconstrained model
will have modified statistics:
\begin{eqnarray}
\label{correctionEqn}
\avg{x_i}_{\alpha = 0} &=& ( 1 - f_0 - f_1 ) \avg{x_i} 
                            + f_{i1} \\
\avg{x_i x_j}_{\alpha = 0} &=& ( 1 - f_0 - f_1 ) \avg{x_i x_j},
\nonumber
\end{eqnarray}
where $f_0$ is the frequency of size zero fights,
$f_{i1}$ is the frequency of size one fights consisting
solely of individual $i$, and
$f_1 = \sum{f_{i1}}$ is the overall frequency of
size one fights (all measured in the unconstrained model).
In terms of unconstrained individual frequencies, these are
\begin{eqnarray}
f_0 &=& \prod_i (1 - \avg{x_i}_{\alpha = 0}) \\
f_{i1} &=& \avg{x_i}_{\alpha = 0} \frac{f_0}{1 - \avg{x_i}_{\alpha = 0}}.
\end{eqnarray}
We use an iterative
procedure to solve equations~(\ref{correctionEqn})
for $\avg{x_i}_{\alpha = 0}$, which are then used in \eqref{fieldsEqn}
to find the fields.  Finally, samples from the independent model [\eqref{constrainedIndependentEnergy}] are produced by sampling using \eqref{unconstrainedIndependentEnergy} and simply discarding samples in which fewer than 2 individuals appear.  For our data, this results in discarding about 17\% of samples produced with \eqref{unconstrainedIndependentEnergy}.

\subsection{Pairwise maximum entropy model inference}
\label{IsingModelSection}

We next constrain our equilibrium maximum entropy model to match the frequencies of appearance of both individuals and pairs of individuals. This model is known to be the spin-glass Ising model \cite{SchBerSeg06}, with relative negative log-likelihood
\begin{equation}
L^{\mathrm{SG}}_{\alpha = 0}(x) = \sum_{ij} - x_i J_{ij} x_{j}.
\end{equation}

As in the independent model, the likelihood of fights with fewer than two individuals is taken to zero:
\begin{equation}
\label{IsingE}
L^{\mathrm{SG}}_{\alpha \rightarrow \infty}(x) = \lim_{\alpha \rightarrow \infty}
\sum_{ij} - x_i J_{ij} x_{j} + \alpha ~ \Theta \left( 2 - \sum_k x_k \right).
\end{equation}

% see notes 2/21/2014 2, 2/24/2014 2, 2/25/2014 2

The statistics we fit in the equilibrium model are the individual and pairwise frequencies of appearance:
\begin{eqnarray}
f_{i} &=& \avg{x_i} = \frac{N(x_i)}{N}, \\
f_{ij} &=& \avg{x_i x_j} = \frac{N(x_i,x_j)}{N} ~~~~ \mathrm{for} ~ i \neq j,
\end{eqnarray}
with $N(x_i)$ and $N(x_i,x_j)$ representing, respectively, the observed number of appearances in unique fights of the individual $i$ and the pair $i,j$.

We would like to fit the individual and pairwise frequencies of appearance to the precision that the data supports.  As a measure of the goodness of fit, we use the average of normalized residuals
\begin{eqnarray}
\label{isingCost}
\avg{\chi^2_{\mathrm{SG}}} &=& \frac{2}{n(n+1)} \sum_{i} \sum_{j \geq i}
\frac{ (f_{ij}^{\mathrm{data}} - f_{ij}^{\mathrm{model}})^2 }
{ \sigma_{f_{ij}}^2 },
\end{eqnarray}
where
\begin{equation}
\sigma_{f_{ij}}^2 = \frac{ f_{ij} (1-f_{ij}) }{ N }
\end{equation}
is the expected variance of each residual due to finite $N$,
and repeated indices are understood as individual frequencies: $f_{ii} = f_i$.  A good fit is expected to have $\avg{\chi^2} \approx 1$.

To perform fitting, we use a simplified method that starts with the mean field solution and varies a single parameter corresponding to weighting a non-interacting prior. Specifically, we make use of the $L_2$-regularized mean field entropy of Ref.~\cite{BarCoc13}.  The regularization consists of a Gaussian prior with a form designed to make the mean-field case easily solvable, corresponding to an additional term in the relative negative log likelihood
\begin{equation}
L_{\mathrm{prior}} = \gamma \sum_i \sum_{j > i} J_{ij}^2 f_i(1-f_i)f_j(1-f_j),
\end{equation}
where $\gamma$ is the strength of the prior, which favors interactions
$J_{ij}$ that are smaller in magnitude.
The mean field solution under this regularization is \cite{BarCoc13}
\begin{eqnarray} \nonumber
J^{\mathrm{MF}}_{ij} &=& \frac{J'_{ij}}
{\sqrt{f_i(1-f_i)f_j(1-f_j)}}, ~~ i \neq j \\
J^{\mathrm{MF}}_{ii} &=& \sum_{j\neq i} J^{\mathrm{MF}}_{ij} \left(
(f_{ij} - f_i f_j)\frac{f_i - \frac{1}{2}} {f_i(1-f_i)} - f_j
\right),
\label{JMF}
\end{eqnarray}
where $J'$ is the matrix that has the same eigenvectors $v_q$
as the correlation matrix
\begin{equation}
C = \frac{f_{ij} - f_i f_j}{\sqrt{f_i(1-f_i)f_j(1-f_j)}}
\end{equation}
and eigenvalues $j_q = 1/\hat c_q$, where $\hat c_q$ are
regularized versions of $C$'s eigenvalues $c_q$:
\begin{equation}
\hat c_q = \frac{1}{2} \left( c_q - \gamma +
\sqrt{(c_q - \gamma)^2 + 4 \gamma} \right).
\end{equation}

This regularization is typically used in a Bayesian sense to avoid overfitting, where a typical value of the regularization strength is $\gamma = 1/(10 N f^2 (1-f)^2)$, with $N$ the number of samples and $f = n^{-1} \sum_i f_i$ the average individual frequency \cite{BarCoc13}.  Avoiding overfitting, however, still typically requires either enormous $N$ (e.g.~\cite{SchBerSeg06}) or restricting the effective number of fit parameters via an expansion (e.g.~\cite{BarCoc13}).  In our case, $N$ is fundamentally limited in that we are describing a stable social epoch of finite duration.  In addition, typical high-temperature expansions cannot easily incorporate the restriction that fights have a minimum number of participants [the $\alpha$ term in \eqref{IsingE}].

% Could say something here about the typical size of the fit gammaprime
% (where gamma = gammaprime / (N f^2 (1-f)^2) and BarCoc13 used
% gammaprime = 0.1)
% Though this seems to depend somewhat on N, we are getting
% gammaprime ~ 5 or 10.  See stdout_stderr_0182.txt, etc.

Alternatively one can treat $\gamma$ as a fitting parameter that interpolates between the mean-field solution (which we find overestimates the strength of interactions) and the case of independent individuals. Although it is not \textit{a priori} obvious that varying this single parameter will be enough to fit the observed statistics within expected statistical fluctuations, we find that this is true for our data.  Numerically sampling from the distribution defined by \eqref{IsingE} with the regularized mean-field $J$ from \eqref{JMF}, we minimize $\avg{\chi^2}$ from \eqref{isingCost} as a function of $\gamma$.  Sampling is performed using a standard Metropolis Markov Chain Monte Carlo approach.  In evaluating the fit, we choose the number of samples to scale with the number of data samples, using $N_\mathrm{samples} = 20 N$.

\begin{figure}
\centering
\includegraphics[scale=0.80]{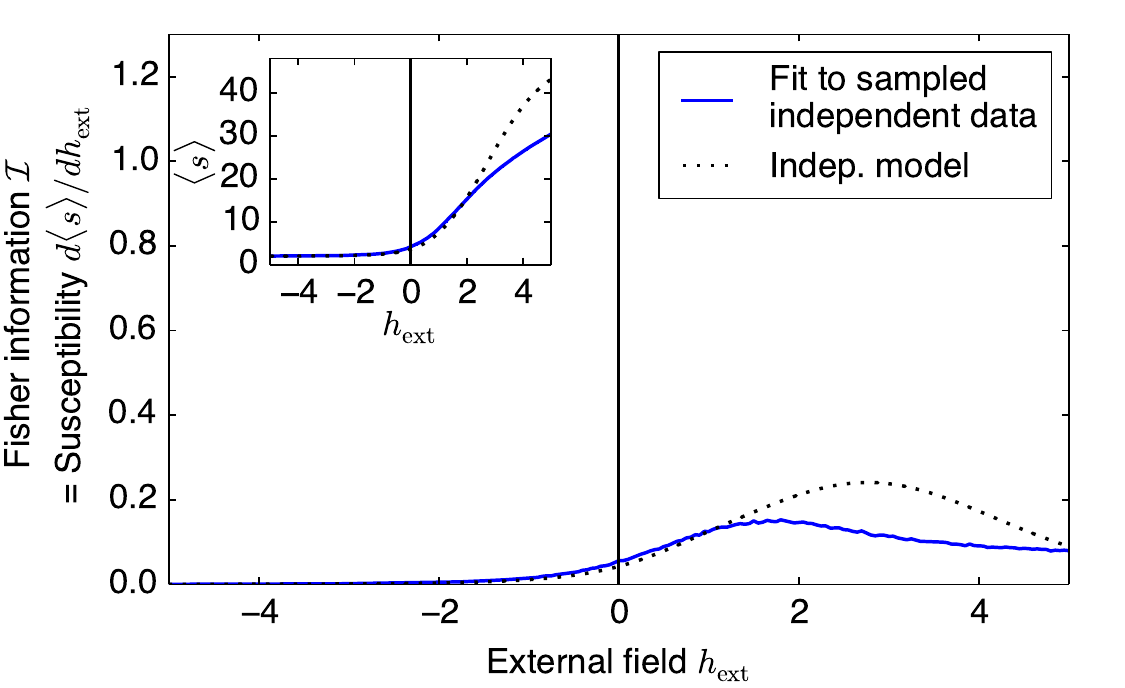}
\caption{
Performing the second-order maximum entropy fitting procedure
on an amount of data equal to the observed fights drawn from
a noninteracting model produces a flat susceptibility curve comparable to the
known exact susceptibility for this case.
\label{fitJFreq}%
}
\end{figure}

As a simple check that this inference approach is not biased to find more sensitive systems, we infer a pairwise model using data produced by the independent model and compare its susceptibility to the known exact value [\eqref{suscFreq}] in \figref{fitJFreq}.  The resulting inferred model has susceptibility that stays close to the true value for small $\hext$ and remains smaller than the true value for larger $\hext$.

\subsection{Branching process inference}
\label{BranchingProcessSection}

In the branching process model, we use time-ordered appearance data, fitting the time-ordered conditional appearance probabilities
\begin{equation}
P_{ij} = \frac{ N(x_i^t, x_j^T) }{ N(x_i) },
\end{equation}
where $N(x_i^t, x_j^T)$ counts the number of times individual $j$ appears in the same fight bout as individual $i$, but at a later time $T > t$.\footnote{Note that this is different than previous work in Inductive Game Theory \cite{DeDKraFla10}:
there (time directed) correlations were measured between individual appearances in separate fight bouts, whereas here we measure correlations within fight bouts.}

The parameters we vary in the heterogeneous branching process model are the single-step conditional probabilities $p_{ij}$, which measure the probability that individual $j$ appears in step $t+1$ of the branching process given that individual $i$ appeared in step $t$. Being probabilities, $p_{ij}$ are constrained to values $0 \leq p_{ij} \leq 1$.  (Note that the branching model is thus limited in the extent to which it can represent inhibitory interactions, \emph{e.g.} if $i$'s involvement in the fight deters $j$ from joining.)  We modify this constrained optimization problem into an unconstrained one by defining $\bar p_{ij} = | \tanh^{-1} p_{ij} |$ and performing the optimization over the (unconstrained) $\bar p_{ij}$ parameters.

In the branching process simulation, the first individual to join each fight bout is chosen randomly with probability proportional to the frequency with which each individual appears at the beginning of fights in the data.  At each subsequent time step in the branching process, each individual $j$ who has not yet been activated in the current fight has a probability of joining equal to the sum of $p_{ij}$ for all $i$ active in the previous time step.  The fight bout concludes when no individuals are active in a given time step.  As discussed above, fight bouts that do not grow beyond a single individual are discarded.

Analogously to the case of the equilibrium maximum entropy model, the branching process parameters are fit by minimizing
\begin{equation}
\label{branchingCost}
\avg{\chi^2_{\mathrm{Branching}}} = \frac{1}{n(n-1)} \sum_i \sum_{j \neq i} \frac{ \left(
P_{ij}^{\mathrm{data}}
- P_{ij}^{\mathrm{model}} \right)^2 }{ \sigma_{P_{ij}}^2 },
\end{equation}
where
\begin{equation}
\sigma_{P_{ij}}^2 = \frac{ P_{ij} (1-P_{ij}) }{ N(x_i) }.
\end{equation}
%and the mean Bayesian (Laplace's method)
%\begin{equation}
%P_{ij}^B = \frac{ N(x_i^t, x_j^T) + 1 }{ N(x_i) + 2 }
%\end{equation}
We use a standard Levenberg-Marquardt algorithm ({\tt scipy.optimize.leastsq}) that uses individual residuals and a lowest-order approximation for the Jacobian with respect to parameters. We find that restarting the Levenberg-Marquardt routine every 10 steps (effectively resetting its damping parameter to avoid unnecessarily small steps arising from its assumption of a non-stochastic objective function) produces faster convergence with fewer samples required for each estimate of the residuals and Jacobian. Minimization is stopped once $\avg{\chi^2}$ defined in \eqref{branchingCost} falls below 1.  In addition, we find that switching between simple gradient descent steps and Levenberg-Marquardt minimizations allows for more efficient fitting of larger $P_{ij}$.

\begin{figure}
\centering
\includegraphics[scale=0.75]{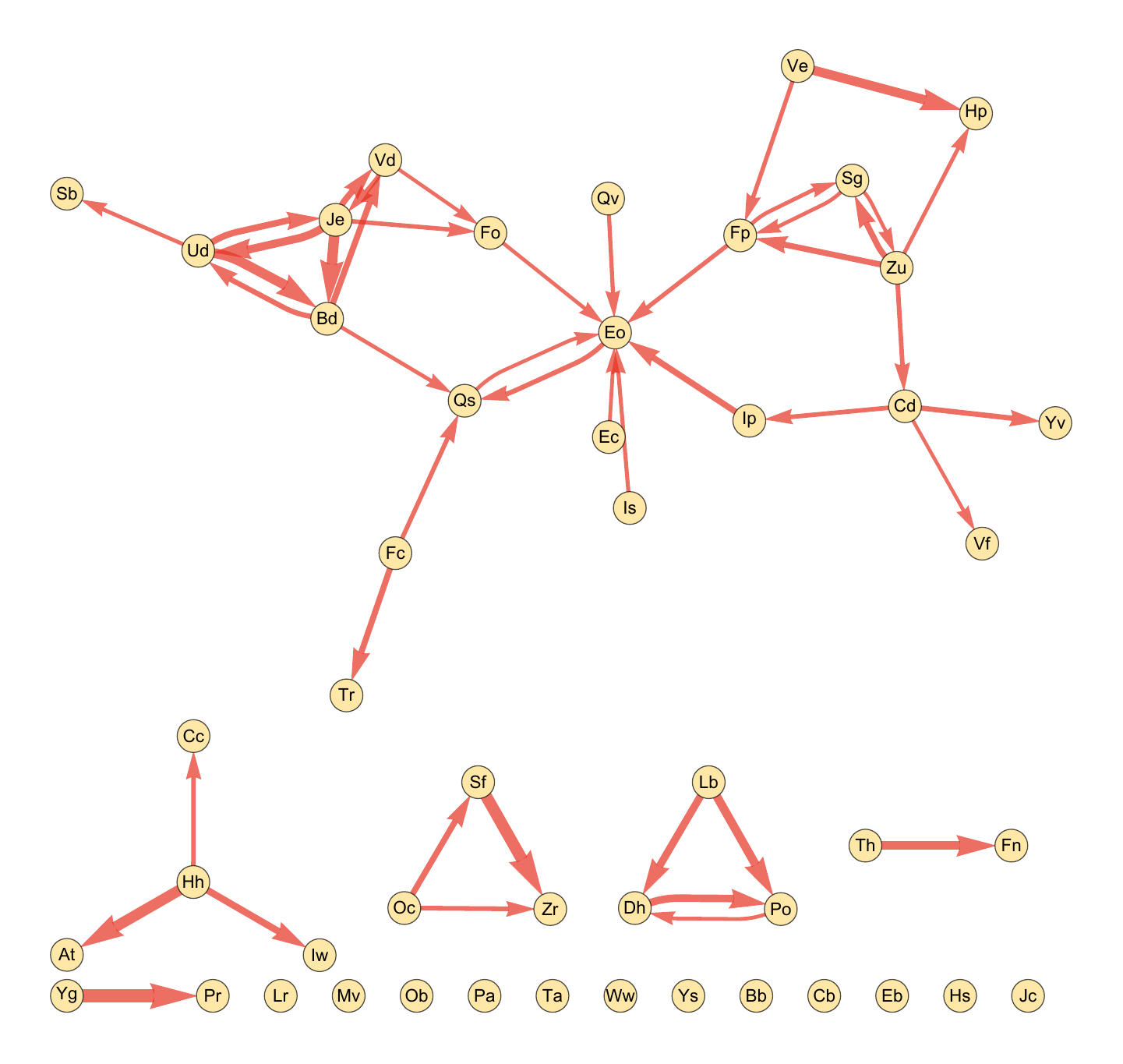}
\caption{ A visualization of the inferred branching process graph.
In the full branching process model, all possible directed pairwise
interactions are included, but here we visualize the most important
interactions by displaying those with triggering probabilities $> 0.09$.
The thickness of each arrow corresponds to the probability of
triggering, with the thickest
lines correspond to a probability of about 1/3: for instance, when Yg appears, Pr
will be triggered by Yg to join with probability of about 1/3.
\label{branchingGraph}%
}
\end{figure}

The resulting inferred redirection probabilities $p_{ij}$ are visualized in \figref{branchingGraph}.

\section{Notes on model evaluation}
\label{modelevaluation}

\begin{figure}
\centering
\includegraphics[scale=0.6]{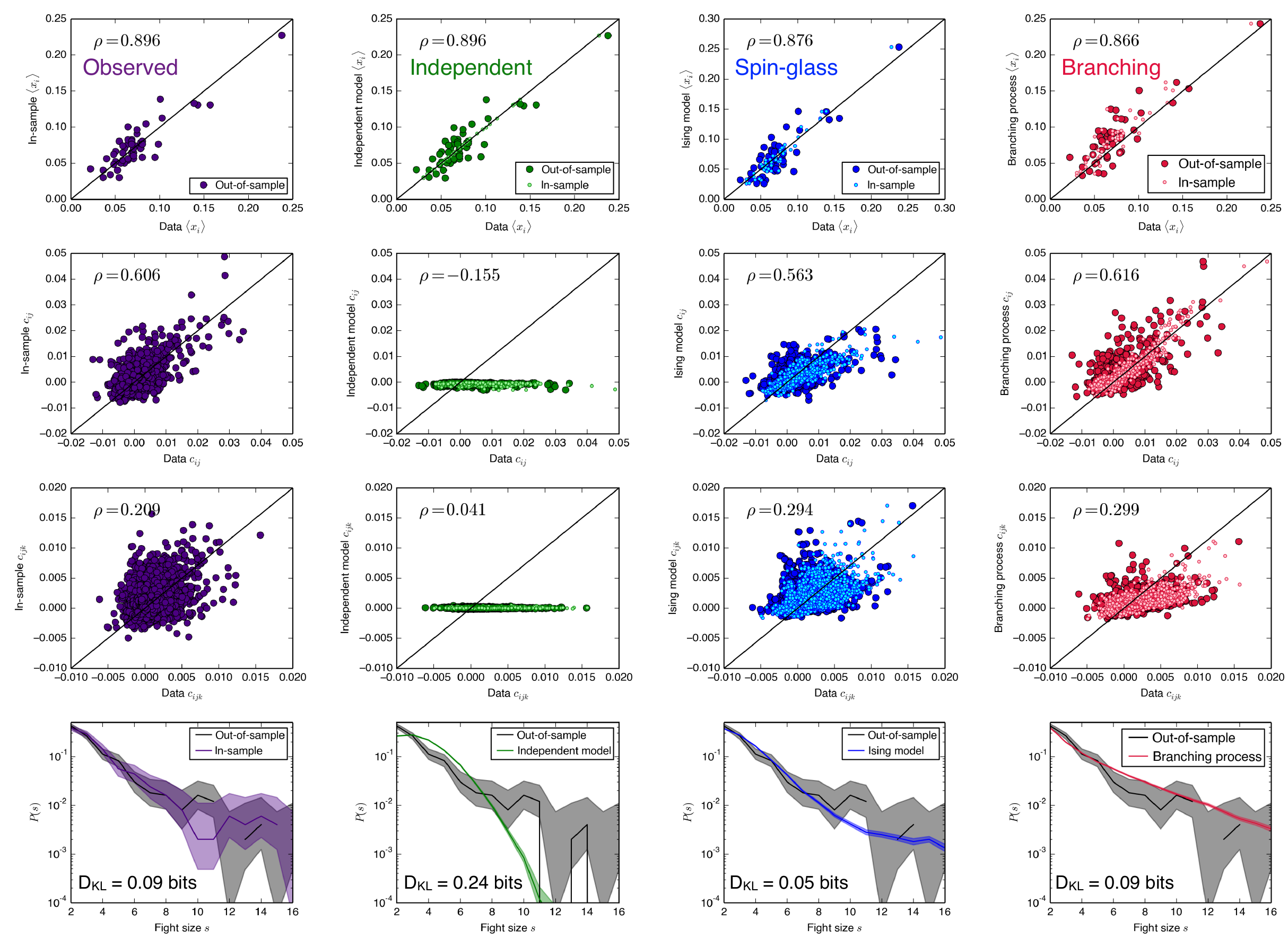}
\caption{
The degree of fit for the noninteracting model (green),
maximum entropy pairwise model (blue), and
branching process model (red) to out-of-sample data, compared
to the same for the in-sample data (indigo) to which the models are fit.
For each model, $10^5$ samples were taken to evaluate predicted
statistics.  Also shown on each plot is the Pearson correlation $\rho$ between predicted and out-of-sample statistics (for individual, pairwise, and triplet statistics) or the Kullback-Leibler divergence $D_{\mathrm{KL}}$ between predicted and out-of-sample distributions (for fight sizes).  (To avoid problems with large fight sizes that are never observed, $D_{\mathrm{KL}}$ is calculated only using fights of size $\leq 12$.)
\label{fittingScorecardFigure}%
}
\end{figure}

To check the performance of each of our models, we first compare statistics computed with the model to those computed on out-of-sample data.  The results for a single choice of in-sample data are shown in Fig.~\ref{fittingScorecardFigure}.  Half of the fights are randomly chosen as in-sample data, with the remaining treated as out-of-sample data to be predicted.  We see that the independent model does not capture second- or third-order statistics nor the distribution of fight sizes, while both the equilibrium and dynamic models capture these to produce predictions that are roughly as accurate as using out-of-sample data.

\begin{figure}
\centering
\includegraphics[scale=0.80]{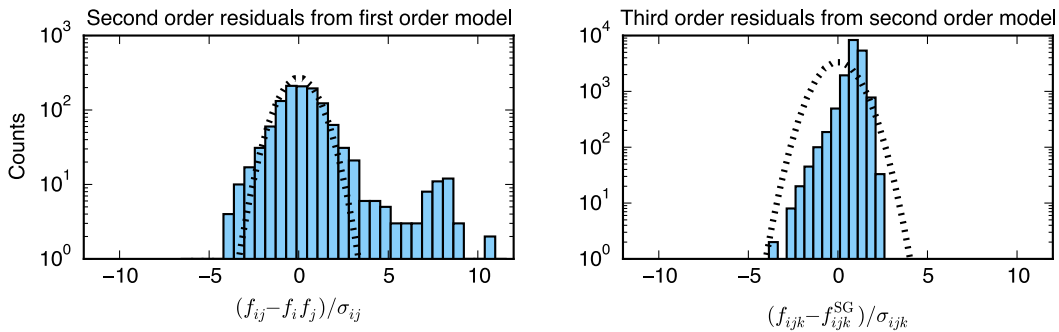}
\caption{
Second- and third-order statistics in the conflict data.  Second-order
statistics (left) clearly violate the null expectation for a
first-order independent model (dotted line), while third-order statistics (right)
lie within expected fluctuations from a second-order model (dotted line).
$f_{ijk}$ is the empirical frequency with which each triplet appears in fights,
$f^\mathrm{SG}_{ijk}$ is this frequency in the pairwise equilibrium model, and
$\sigma_{ijk} = \sqrt{f_{ijk}(1-f_{ijk})/N}$ is the expected standard deviation.
\label{isingStatistics}%
}
\end{figure}

Second, we can check that residuals lie within the bounds of expected statistical fluctuations from finite sampling.  Shown in Table~\ref{goodnessOfFitTable}, the equilibrium and dynamic models have squared residuals that are below but near the expected value $\avg \chi^2 = 1$, whereas the independent model is inadequate to describe the statistics.  This is visualized in more detail with the distribution of residuals in \figref{isingStatistics}.

\begin{table}
\begin{center}
\caption{
\label{goodnessOfFitTable}
Goodness of fit to data for the three models,
calculated using \eqref{isingCost} for the independent and pairwise maximum entropy models and \eqref{branchingCost} for the dynamic branching model.
With $\avg{\chi^2} \sim 1$, the equilibrium pairwise and dynamic branching models fit the data roughly within the precision afforded by the data.
Overfitting, which would be indicated by $\avg{\chi^2} \ll 1$, is avoided by
using constrained minimization in the case of the spin-glass model
(see Sec.~\ref{IsingModelSection}) and by ending minimization once
$\avg{\chi^2} \leq 1$ in the case of the branching model
(see Sec.~\ref{BranchingProcessSection}).}
\begin{tabular}{  l | l | l | l }
& \emph{Independent model}
& \emph{Equilibrium pairwise model}
& \emph{Dynamic branching model}                        \\
\hline
Random half of data
&  $\avg{\chi^2} = 1.134 $
&  $\avg{\chi^2} = 0.459 $
&  $\avg{\chi^2} = 0.414 $                              \\
All data
&  $\avg{\chi^2} = 1.822 $
&  $\avg{\chi^2} = 0.597 $
&  $\avg{\chi^2} = 0.462 $                              \\
\end{tabular}
\end{center}
\end{table}

We find no evidence of significant higher order correlations in the data (\figref{isingStatistics}) and we therefore do not explore models with interactions of higher order. We note however that the resolution of higher order correlations is limited by the finite number of observed fights and relatively small frequency of individual participation. This cannot easily be remedied by collecting more data as the system is not at equilibrium over longer timescales. To deal with this, we must restrict the data we use in the analyses to collection windows defined by ``socially stable periods'' (SI, Sec~\ref{EmpiricalMethods}).

\section{Notes on evaluating sensitivity and stability}
\label{evalsens}
Phase transitions are typically identified as conditions under which the varying of a control parameter causes large-scale changes in the behavior of a system, in a way that sensitivity per individual (measured by, \emph{e.g.}, specific heat or susceptibility) grows arbitrarily large with growing system size. This becomes possible only when there is a collective instability, meaning that the effective size of the perturbation (that starts, say, with a single individual) does not shrink as it spreads through the system but stays of constant size or grows (potentially affecting all individuals).  Thus in a finite system, the combination of a peak in sensitivity and collective instability can be used as an indicator of a phase-transition-like state.

\subsection{Sensitivity as Fisher information}
In our finite system the notion of diverging sensitivity is arguably more accurately described in terms of information theory. Even when the idea of a phase transition becomes fuzzy in a finite system, the Fisher information measures something adaptively important: the degree to which individual scale perturbations are visible at the global scale, or, equivalently,  the connection between the behavior of any individual and the behavior of the whole \cite{Tchernookov2012,Prokopenko2011}.

\subsection{Analytical results for sensitivity in the independent model}

Here we show that the sensitivity (susceptibility) to increased aggression in the independent model can be efficiently solved numerically.  This is used to make a comparison with the pairwise equilibrium model in \figref{susceptibilityFigure}.

First, in the more analytically straightforward case in which we allow fights of size zero and one ($\alpha = 0$; see section \ref{independentInference}), the average fight size and susceptibility are
\begin{eqnarray}
\avg{s}_{\alpha = 0} &=&
\sum_i \left(1 + \exp( h_i - \hext ) \right)^{-1}               \\
\chi_0 = \frac{\partial \avg{s}_{\alpha = 0} }{\partial \hext}
&=& \sum_i \sech^2 (h_i - \hext).
\end{eqnarray}
The partition functions of the constrained and unconstrained models,
defined such that
\begin{eqnarray}
p(\vec x)_{\alpha = 0} &=& \exp[ - L_{\alpha = 0}(\vec x) ] / Z_0  \\
p(\vec x)_{\alpha \rightarrow \infty} &=&
 \exp[ -L_{\alpha \rightarrow \infty}(\vec x) ] / Z_\infty,
\end{eqnarray}
are given by
\begin{eqnarray}
Z_0 &=& \prod_i (1 + \exp h_i) \\
Z_\infty &=& Z_0 - 1 - \sum_i \exp h_i.
\end{eqnarray}
In terms of these values, when fights of size zero and one are forbidden
($\alpha \rightarrow \infty$), the average fight size and susceptibility become
\begin{eqnarray}
\avg{s}_{\alpha \rightarrow \infty} &=&
\frac{Z_0}{Z_\infty} \avg{s}_{\alpha = 0}
- \frac{\sum_i \exp h_i}{Z_\infty}                               \\
\chi_\infty = \frac{\partial \avg{s}_{\alpha \rightarrow \infty} }
{\partial \hext}
&=& \frac{Z_0}{Z_\infty}
\frac{\partial \avg{s}_{\alpha = 0} }{\partial \hext}
- \frac{\sum_i \exp h_i}{Z_\infty}
+ \frac{Z_0}{Z_\infty} \avg{s}_{\alpha = 0}^2
- \avg{s}_{\alpha \rightarrow \infty}^2. \label{suscFreq}
\end{eqnarray}

\subsection{Operationalizing collective instability in the branching process model}
\label{collectinstab}
Collective instability is straightforward to understand in a branching model. In an infinite system, this model has a well-defined phase transition when $R_0$, the average number of other individuals triggered when a single individual becomes active, is equal to 1. This also corresponds to the point at which the system is maximally sensitive to changes in parameters. The fact that this ``local'' amplification factor is also indicative of a global transition relies on the infinite limit:  In a finite system, cascades will be shortened when they reach individuals that have already been activated, and maximal sensitivity will happen at some $R_0 > 1$ \cite{Beg07}.  We thus think of $R_0$ as measuring a ``local'' or lowest-order stability.
In the heterogeneous branching process model, this linear stability of the peaceful state is indicated by the largest eigenvalue $R_0$ of the redirection probability matrix $p_{ij}$.

We note that in the branching model, as opposed to the equilibrium model, increasing activation will always eventually lead to instability.

\subsection{Operationalizing collective instability in the pairwise equilibrium model}
\label{stabilitySection}

In an infinite system, the pairwise equilibrium model also has a phase transition under the condition of local instability, with a corresponding diverging sensitivity.  In this case, instability can be quantified using the mean-field solution, connecting with a high-temperature expansion of spin-glass models.

One way to think about the continuous phase transition in an infinite spin-glass model is that it is the point at which the high temperature mean-field solution becomes unstable. Mean-field solutions are characterized by frequencies $\vec f$ of individual appearance ($f_i = \avg{x_i}$) that satisfy the self-consistency equation \cite{Sta71}
\begin{equation}
\label{meanFieldEqn}
f_i = F_i(\vec f) \equiv
\left[ 1 + \exp \left(J_{ii} + 2 \sum_{j \neq i} J_{ij} f_j\right) \right]^{-1}.
\end{equation}
Intuitively, individual $i$'s frequency of fighting is determined by the mean ``field'' it feels as a result of others' fighting.  The function $F_i$ encodes how $i$ reacts to its environment, translating the mean fighting frequencies $i$ sees into its own mean frequency.  When \eqref{meanFieldEqn} holds for every individual using a single set of frequencies $\vec f$, this defines the mean field solution.

Now imagine perturbing fighting frequencies $\vec f$ by a small $\Delta \vec f$.  This will typically no longer be a solution of \eqref{meanFieldEqn}.  But if we repeatedly apply the function $F$ to $\vec f + \Delta \vec f$, we can imagine two possibilities: we might end up back at $\vec f$ (so that $\lim_{n\rightarrow \infty} F^n(\vec f + \Delta \vec f) = \vec f$), or we might get further and further from $\vec f$.  We will call the first case a ``stable'' mean field solution and the second case ``unstable''.

For small perturbations $\Delta \vec f$, we can distinguish these two cases
by taking a derivative to perform a linear stability analysis. Specifically,
\begin{equation}
F_i(\vec f + \Delta \vec f) \approx F_i(\vec f)
+ \sum_j \frac{\partial F_i}{\partial f_j} \Delta f_j,
\end{equation}
and to converge back to $\vec f$ for every perturbation, we must have that the updated perturbation along each direction
has shrunk.  This corresponds to a condition on the eigenvalues
$\lambda_\alpha$
of the derivative matrix $M_{ij} \equiv \partial F_i / \partial f_j$;
the state is stable if
\begin{equation}
| \lambda_\alpha | < 1 ~~~~ \forall ~ \alpha.
\end{equation}
Thus the eigenvalue $\lambda$ with largest magnitude determines stability.

We can write this derivative matrix $M$ more explicitly by taking the
derivative of \eqref{meanFieldEqn} and assuming that we are at the
fixed point $(\vec f = F(\vec f))$:
\begin{eqnarray} \nonumber
M_{ij} &=& \frac{\partial F_i}{\partial f_j} \\ \nonumber
&=& -2 (1 - \delta_{ij}) J_{ij}
\exp \left(h_i + 2 \sum_{j \neq i} J_{ij} f_j\right) f_i^2 \\ \nonumber
&=& -2 (1 - \delta_{ij}) J_{ij} \frac{1 - f_i}{f_i} f_i^2 \\
&=& -2 (1 - \delta_{ij}) J_{ij} f_i (1 - f_i).
\label{Meqn}
\end{eqnarray}

Thus $M$ is a matrix analogous to $p_{ij}$ in the branching process model in that its spectrum is informative about how perturbations grow or shrink. Specifically, we use the magnitude $\lambda$ of the largest eigenvalue of $M$ as a measure of stability of the system.  When $\lambda > 1$, we expect the system to be unstable to perturbation.

This condition on the stability of mean field theory can be shown to be equivalent to the condition that identifies the spin-glass transition in an infinite system.
%An infinite spin-glass system with heterogeneous $J_{ij}$ coupling
%all individuals (known as the Sherrington-Kirkpatrick model
%when $J_{ij}$ are chosen from a Gaussian) has interesting behavior
%only when $J_{ij}$ scales as $1/N$.  This means that an expansion
%that treats $J_{ij}$ is small should be sufficient
Specifically, instability of the high-temperature mean-field expansion happens only below the spin-glass temperature \cite{MezParVir87,Geo04}. To lowest order in $1/T$ (corresponding to lowest order in $J_{ij}$ or $1/N$), the mean-field free energy has the form (following \cite{GeoYed91})
\begin{equation}
A = \sum_i f_i \log f_i + (1-f_i) \log (1-f_i) - \sum_i \sum_{j \neq i} J_{ij} f_i f_j - \sum_i J_{ii} f_i,
\end{equation}
which when differentiated produces the self-consistency equation
(\ref{meanFieldEqn}).  Taking a second derivative,
\begin{equation}
\frac{\partial^2 A}{\partial m_i \partial m_j} =
    \delta_{ij} \frac{1}{f_i(1-f_i)}
    - (1 - \delta_{ij}) 2 J_{ij} \equiv \Lambda_{ij},
\end{equation}
which defines stability to this order when all eigenvalues of $\Lambda_{ij}$ are positive \cite{Geo04}. Because $f_i(1-f_i)>0$, this condition is the same as all eigenvalues of $f_i(1-f_i)\Lambda_{ij} = \delta_{ij} - M_{ij}$ being positive (where $M$ is defined in \eqref{Meqn}), which is in turn equivalent to the above condition that all eigenvalues of $M_{ij}$ are less than 1.

To create a homogeneous finite system poised at the transition defined by the eigenvalue $\lambda$ (shown as the red dotted curve in \figref{susceptibilityFigure}), we define a homogeneous positive $h$ and negative $J$ that make each individual's frequency $f = 1/2$ and $\lambda = 1$.  
%\textit{BCD: Are all eigenvalues of $M$ equal to 1 in this case?}

\section{Notes on quantifying DFC}
\label{sensitiveindividuals}
As described in the main text, by simulating forcing some number of individuals to be involved in every fight we can quantify distance from the critical point by measuring changes to sensitivity and collective instability.  We measure this distance in units natural to the system by forcing a subset of individuals to join fights and measuring the sensitivity of the remaining individuals.

The number of individuals (or, in principle, subgroups) that must be forced to reach peak sensitivity provides an operational definition of DFC. We find three to five individuals is sufficient and we find individual variation in sensitivity scores. In both the equilibrium and branching process models individuals with higher sensitivity scores are those who (by definition) exert greater influence on how far the system sits from the critical point.

We find that individuals whose simulated forcing causes the largest increase in average fight sizes are also those who cause the largest increase in sensitivity, as might be expected in this system with largely excitatory interactions.  In each model, this sensitivity score depends on the individual's place within the overall network structure, measuring both its direct influence on inducing others to fight and indirect influence through those it induces to fight.

%ALSO ADD COMMENT ON COUZIN 6 --number of individuals that need to be informed for the group to make a coherent decision.

\subsection{Tuning mechanisms}
\label{tuning}
The sensitivity scores generated by each model are capturing different underlying tuning mechanisms. In the case of the branching process model, the model captures the spread of fight-joining through direct interactions: one individual through its behavior triggers the involvement of a second individual. Hence this model only allows that the target individuals, through their own fight joining decisions, can up or down regulate their fight involvement. When a high sensitivity individual up or down regulates its own behavior moving the system respectively closer to or further from the critical point, we call this \emph{direct tuning}.

The equilibrium model, on the other hand, is agnostic to cause, recording any pairwise correlation in fight joining regardless of which individual triggered the joining event. As such it captures the full space of behavioral mechanisms leading to fight-joining---individuals can become involved in fights by up regulating their own fight involvement and \emph{also} through changes to third-party behavior. For example, policing (by third-parties to the fight) and other conflict management mechanisms reduce the frequency of redirected and directed aggression in the system, and as such can dampen the fight joining behavior of the target individual. When a third party, like a policer, up or down regulates the behavior of high sensitivity individuals we call this \emph{indirect tuning}.

\subsection{Tune towards robustness or towards criticality?}
\label{adaptive value}
A natural question raised by the discoveries that DFC can be tuned and that there are behavioral mechanisms in place that in principle allow this tuning is whether to tune towards robustness (increase DFC) or criticality (decrease DFC). This decision depends on three factors: (1) the adaptive utility of criticality--when does it make sense to sit near the critical point?, (2) the true state of the environment, and (3) the perceived state of the environment---the accuracy with which system components can detect and encode the environmental state.

The consequences of being near or at the critical point is that information can propagate quickly, with small changes to component behavior inducing large-scale changes in both structure and function. Hence criticality allows the system to more easily reconfigure. How does this work? Moving towards criticality changes the distribution of fight sizes such that there are more large fights. Reconfiguration becomes more likely with large fights because large fights cost more\cite{Dedeo2010} and costly conflict can lead to changes in alliances, coalitions and the power structure, which controls the cost of conflict management\cite{Flack:2006jh,Brush2013}. 

We predict that reconfiguration is adaptive when the environment, after having been stable for some time, becomes uncertain. Hence when the state of the environment is stable or very slowly changing and represented by a delta function, the optimal choice (we propose) is to increase DFC. When the environment is uncertain--that is, when it is not clear what the distribution of environmental states is, the system should decrease DFC.

For tuning to be adaptive, the state of the environment must also be correctly perceived by the tuning agent(s). To make this clear, consider the following: a flock of birds in search of food with high and low sensitivity individuals. Some individuals in the flock have poor eyesight and so misjudge the location of food and others have good eyesight. If the high sensitivity birds are also those with poor eyesight, the system may be inappropriately driven away from food sources. If on the other hand the good eyesight birds are also highly sensitive, the system will be appropriately driven towards food. In a similar way, tuning DFC will require accurate perception of the most beneficial change. \emph{Hence in order for DFC to be tunable and for that tuning to be adaptive in a biological system these two types of heterogeneity must be aligned: high sensitivity score individuals must also be good detectors of environmental state}. Neither of these types of heterogeneity has received much attention in the collective behavior, criticality, or biological phase transition literatures. Consequently little is known about how common it is for these types of heterogeneity to be aligned or whether there are in some systems mechanisms to bring them into alignment.

In our model system, both types of heterogeneity appear to be aligned. Results of earlier work\cite{Flack:2005ih,Flack:2006jh}, suggest the individuals with the highest sensitivity scores also are those whose ``health state" signals the state of the larger social environment. The idea is as follows: the ``health state'' of individuals (healthy or compromised) builds up over a history of affiliative and agonistic social interactions with other group members and hence is a slow variable\cite{FlaErwEll13}. The health state of the high sensitivity individuals serves as a proxy for system state because these individuals are among the weakest in the system (lowest 10 percent of the power distribution) and hence register stressful periods more visibly than the other group  members \cite{FlackHammerKrak2012}. 

When these animals are healthy, the system can be said to be in a stable, low variance period. When these individuals are stressed above some baseline level, social dynamics are becoming uncertain. In either case, when aggression during a fight ``reaches them'', they move the system closer to the critical point \emph{because of their position in the aggression network as represented by the spin glass and branching process models}. Aggression is less likely to reach them when the policing mechanism is functional and more likely to reach them when the policing mechanism has been disabled \cite{Flack:2006jh}.

\begin{figure}
\centering
\includegraphics[scale=0.85]{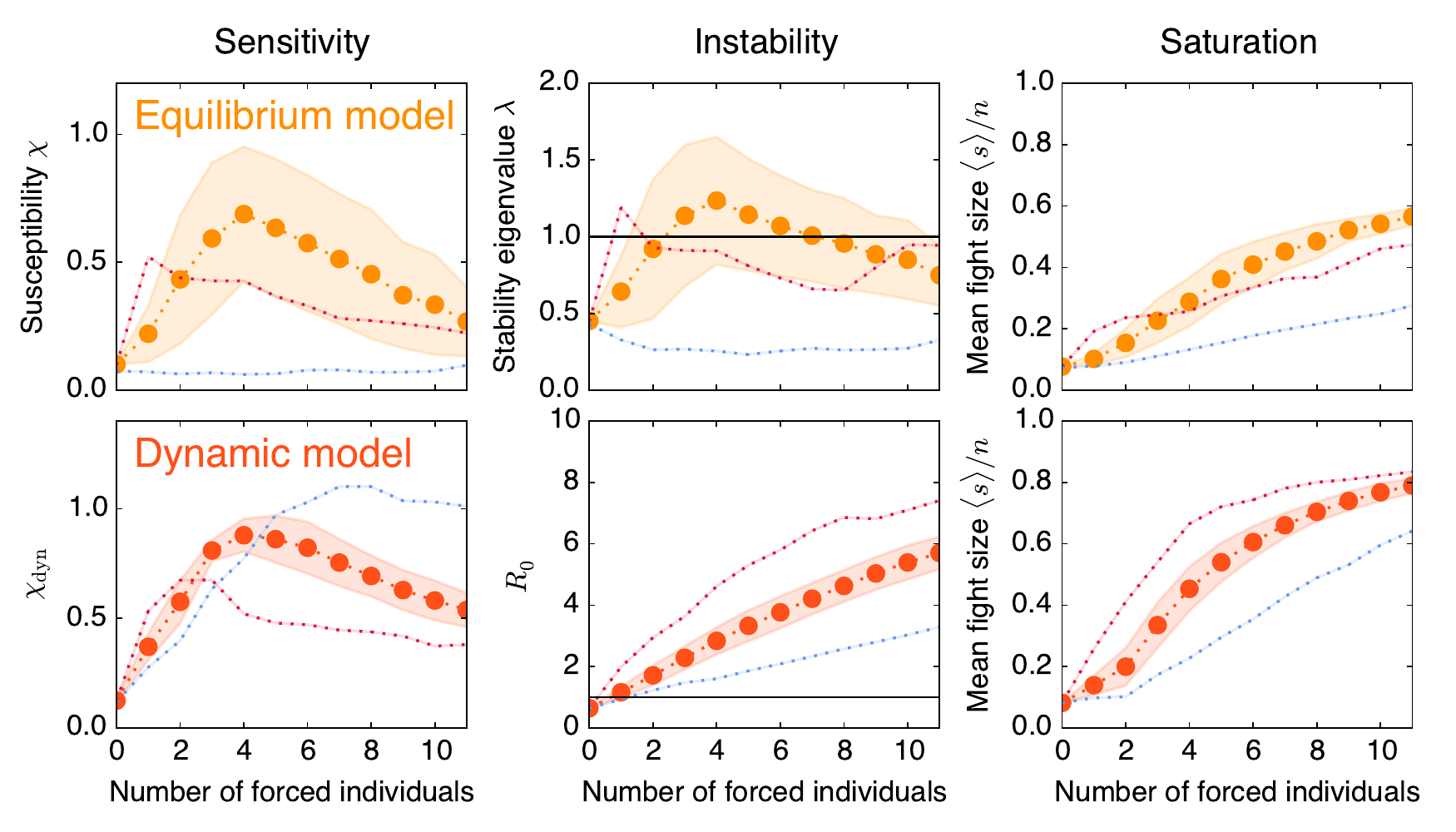}
\caption{ 
Same as \figref{ellStarFigure}, except that the magenta and blue lines here show results when forced individuals are chosen as those with the largest and smallest effect on the mean fight size of remaining individuals when forced in the otherwise unperturbed system.  (\figref{ellStarFigure} reperforms this optimization after adding each individual.)
\label{ellStarFigure2}%
}
\end{figure}

In \figref{ellStarFigure}, the magenta and blue lines demonstrate the potential heterogeneity of responses when different individuals are forced.  The magenta lines demonstrate the effect of forcing individuals in an order that maximizes the resulting average fight size at each step, and blue in an order that minimizes it. The individuals are re-sorted each time another is forced as this can affect the order (for instance, forcing one individual in a strongly-correlated clique can decrease the effect of forcing other individuals in that clique).  In \figref{ellStarFigure2}, we contrast the case in which individuals are sorted by their effect on the original, unperturbed state of the system. The qualitative results are the same as in \figref{ellStarFigure}.

\section{Thermodynamic derivatives and Fisher information in the equilibrium model}
\label{FisherInfoSection}
Quite generally for equilibrium models, it can be shown that the Fisher information is deeply related to important thermodynamic derivatives.
The Fisher information is defined as \cite{CovTho91}
\begin{equation}
\label{FisherEqn}
\F(\mu) = \int
% Manually lowering the subscript for better vertical spacing.
%\bigl(\partial_\mu \log p(x)\bigr)^2 p(x) \, \mathrm{d}x
\left(\frac{\partial \log p(x)}{\partial \mu}\right)^{\raisebox{-2pt}{$\scriptstyle 2$}} p(x) \, \mathrm{d}x,
%\left(\frac{\partial \log p(x)}{\partial \mu}\right)^{2} p(x) \, \mathrm{d}x,
\end{equation}
where $\mu$ parameterizes a distribution $p(x)$ describing the  behavior of a system, and $x$ represents any number of relevant measurable system state variables. Generalized to multiple parameters, the Fisher information matrix is a fundamental object in information geometry, and forms a Riemannian metric that becomes singular precisely at phase transitions \cite{Cro07,Prokopenko2011}. $\F(\mu)$ is typically used to measure the amount of information about $\mu$ that can be inferred from draws from $p(x)$. Conversely, if we view individuals as controlling local parameters $\mu$, $\F(\mu)$ measures the degree of control individuals have on group behavior. Then phase transitions, having diverging $\F$ as $N \rightarrow \infty$, correspond to individuals having arbitrarily large effects. But even at finite $N$, $\F$ measures the amplification of individual information to the global scale. In this sense, $\F$  becomes a straightforward, useful measure of the degree to which a  system's behavior is ``collective.''

Another intuitive meaning comes in terms of the Kullback-Leibler divergence: $\F(\mu)$ represents how quickly the KL-divergence increases as $\mu$ is changed, such that
\begin{equation}
D_\mathrm{KL} \left( p(x|\mu) || p(x|\mu + \Delta \mu) \right)
= \F(\mu) (\Delta \mu)^2 / 2 + O(\Delta \mu^4).
\end{equation}
Thus the Fisher information measures how quickly the modified distribution becomes distinguishable from the original as $\mu$ is varied, and if logs are taken with base 2, $\F(\mu)$ has units of bits per [unit of $\mu$]$^2$.

In the case of an equilibrium system described by a Boltzmann distribution, the Fisher information with respect to a local field $\mu$ is particularly simple, equal to the derivative of the mean of its conjugate variable $x_\mu$, the generalized susceptibility $\mc{I}(\mu) = \frac{\partial}{\partial \mu} \avg{ x_\mu }$.  This example provides a clear link between thermodynamics and information theory.  (Yet the Fisher information  measure is not limited to equilibrium models, generalizing to dynamic out-of-equilibrium systems by simply interpreting $p(x)$ in \eqref{FisherEqn}  as a distribution over relevant output measurements given some known initial conditions.)

This connection between Fisher information and thermodynamic derivatives is well-established \cite{Prokopenko2011}.  Assume we have a system whose distribution over possible states $x$ takes the form of a Boltzmann distribution:
\begin{equation}
p(x) = Z^{-1} e^{- L(x)}.
\end{equation}
Taking a derivative of $\log p(x)$,
\begin{eqnarray}
\frac{ \partial \log p(x) }{ \partial \mu } &=&
- \frac{ \partial L(x) }{ \partial \mu }
+ Z^{-1} \sum_x \frac{ \partial L(x) }{ \partial \mu } e^{- L(x)} \\
&=& \left \langle \frac{ \partial L }{ \partial \mu } \right \rangle
- \frac{ \partial L(x) }{ \partial \mu },
\end{eqnarray}
which, when inserted in \eqref{FisherEqn}, gives
\begin{equation}
\label{Fij2}
\F(\mu) = \left \langle \left( \frac{ \partial L }{ \partial \mu } \right)^2 \right \rangle
- \left \langle \frac{ \partial L }{ \partial \mu } \right \rangle^2.
\end{equation}
This shows that Fisher information is equal to the variance of the derivative of $L$. We can further relate this to thermodynamic derivatives by noting that $L$ is typically linearly dependent on certain ``fields'' (\emph{e.g.} pressure or magnetic field), with derivatives that correspond to measurable macroscopic properties (\emph{e.g.} volume or magnetization).  This linearity allows us to write the Fisher information even more simply:
when
$\langle \partial^2 L / \partial \mu^2 \rangle = 0$,
\begin{equation}
\label{FijSimple}
\F(\mu) = - \frac{ \partial }{ \partial \mu }
\left \langle \frac{ \partial L }{ \partial \mu } \right \rangle.
\end{equation}
(To see this, explicitly take the derivative of the expectation value:
\begin{equation}
\frac{ \partial }{ \partial \mu }
\left \langle \frac{ \partial L }{ \partial \mu } \right \rangle
= \frac{ \partial }{ \partial \mu } \left[ Z^{-1} \sum_x \exp(-L(x))
\frac{ \partial L(x) }{ \partial \mu} \right]
%&=& Z^{-1} \sum_x \exp(-L(x)) \left[ - \frac{ \partial L(x) }{ \partial \theta_i }
%                                   \frac{ \partial L(x) }{ \partial \theta_j }
%    + \left \langle \frac{ \partial L(x) }{ \partial \theta_i } \right \rangle
%        \frac{ \partial L(x) }{ \partial \theta_j }
%    + \frac{ \partial^2 L(x) }{ \partial \theta_i \partial \theta_j }  \right] \\
= - \left \langle \left( \frac{ \partial L }{ \partial \mu } \right)^2 \right \rangle
+ \left \langle \frac{ \partial L }{ \partial \mu } \right \rangle^2
+ \left \langle
\frac{ \partial^2 L }{ \partial \mu^2 } \right \rangle,
\end{equation}
which is equal to $-\F(\mu)$ from Eq.~(\ref{Fij2}) when the last term is zero.)

Connecting this result to our equilibrium model, the susceptibility and specific heat are related to the Fisher information with respect to external field $\hext$ and temperature $T$:
\begin{equation}
\F(\hext) = \frac{\partial \avg{s}}{\partial \hext} = \chi.
\end{equation}
\begin{equation}
\F(1/T) = -\frac{1}{T^2} \frac{\partial \avg{E}}{\partial T} = C_s.
\end{equation}

The amount of change in the entire distribution over fights is expressed in terms of a single ``order parameter,'' the average fight size in the case of varying $\hext$ (and the average energy in the case of varying $1/T$). This implies that if one is trying to infer small changes in the external field $\hext$ by watching the composition of fights, one loses nothing by simply recording the fight sizes.

%\bibliographystyle{unsrt}

% BCD 6.3.2015
% bibliography for only the supplemental material
\putbib
\end{bibunit}

\end{document}